\documentclass[]{article}
\usepackage{graphicx}
\usepackage{amsmath}
\usepackage{xcolor}
\usepackage{authblk}
\usepackage{hyperref}
\title{Decaying vacuum and evolution from early inflation to late acceleration}
\author{Sarath N  \footnote{\href{mailto:sarath@cusat.ac.in}{sarath@cusat.ac.in}}
	and Titus K. Mathew  \footnote{\href{mailto:titus@cusat.ac.in}{titus@cusat.ac.in}}}
\affil[]{\textit{\small Department of Physics, Cochin University of Science and Technology, \\ Kochi, Kerala 682022, India}}
\date{}
\begin{document}

\maketitle

\begin{abstract}
Decaying vacuum models are a class of models that incorporate the vacuum
energy density as a time-evolving entity that has the potential to explain the
entire evolutionary history of the universe in a single framework. A general
solution to the Friedmann equation can be obtained by considering vacuum
energy density as a function of the Hubble parameter. We have obtained the
asymptotic solution by choosing the appropriate equation of state for matter
and radiation. Finite boundaries in the early and late de Sitter epoch could
be defined by considering the evolution of primordial perturbation wave-
length. An epoch invariant number $N_c$ determines the number of 
perturbation modes that cross the Hubble radii during each epoch has been obtained.
\end{abstract}
\section{Introduction}
As per the current understanding, there exist two accelerating epochs in the evolutionary history of the Universe. The first one is the early inflation driven possibly by a scalar field. The second is the late acceleration in the current epoch in which dark energy is the dominant component. The transient stage between these accelerated expansions consists of a radiation-dominated epoch followed by a matter-dominated one \cite{bjorken2003cosmology}. So far, we do not have an entirely consistent theory that explains the sequential emergence of all these epochs. The inflation stage is characterized by a constant Hubble parameter $H_I$. In contrast, the end accelerated epoch is characterized by another constant, say $\Lambda$, called the cosmological constant, extremely small compared to $H_I.$ The standard $\Lambda$CDM model is very successful in explaining the late acceleration  \cite{bahcall1999cosmic, copeland2006dynamics, 1999ApJ...517..565P, 1998AJ....116.1009R}. However it fails to explain the little value of the so-called cosmological constant. The cosmological constant problem and the coincidence problem (why the dark energy density and dark matter density are comparable to each other in the current 
epoch \cite{Papagiannopoulos2020, Sola:2013gha, weinberg1989cosmological, padmanabhan2003cosmological}) indicate that the rigid nature of the cosmological constant in the entire evolutionary history of the Universe may not be satisfactory.  Understanding this issue 
from a fundamental principle is an important challenge in physics.  This leads to the proposal of dynamical dark energy models by many \cite{lima2013expansion, peracaula2018dynamical, gomez2015dynamical,sola2015hints,george2016holographic}, mainly in the late acceleration context.
There are attempts in the literature to connect the early inflation and the late accelerated epoch. For instance, a particular approach based on particle production is presented in reference \cite{Nunes:2016eab}. 
Another exciting work considers the coupling between a decaying vacuum with radiation and matter \cite{Fay:2014fta, tsiapi2019testing, amendola2000coupled, del2008toward, sharov2017new}. A unification of early inflation and late acceleration was proposed using modified gravity theory \cite{PhysRevD.68.123512, nojiri2003modified}.  Recently Joan Sola et al.  \cite{Sola:2015rra}
proposed a novel class of dynamical vacuum energy models which accounts for the 
the radiation dominated era and subsequent matter dominated epoch and also that for the late accelerating phase are solved separately, using different approximated 
forms for Hubble parameter-dependent vacuum energy density.

In the present work, we solve for a most general common solution by using the Hubble parameter-dependent varying vacuum energy, which represents a smooth 
%all the 
transition of the Universe through subsequent stages from the early inflation. So far up to recently, there is no such constructive effort to find a common solution that could represent both the early inflationary solution and the subsequent epochs up to the end de Sitter epoch. We assume vacuum energy density, modeled
as a power series of the Hubble function \cite{Papagiannopoulos2020, shapiro2002scaling, sola2011cosmologies, sola2013cosmological, sola2014vacuum,john2000generalized} and derive a general solution for the Hubble parameter, which allows getting a complete cosmological scenario with a spacetime 
emerging from an initial de Sitter stage, subsequently evolving into the radiation, matter, and dark energy dominated epochs, which ultimately end to another de Sitter epoch. we define finite boundaries for the Universe, with reference to the model proposed by T. Padmanabhan. \cite{Padmanabhan_2012} by considering the evolution of primordial perturbation wavelength.
\section{General Solution and Asymptotic 
	Epochs}
The Friedmann equations which explain the evolution of the Universe can be written as,
\begin{equation}
\label{eqn:h01}
3H^2 = 8\pi G(\rho_m+\rho_{\Lambda}),
\end{equation} 
\begin{equation}
2\dot{H} + 3H^2 = -8\pi G(\omega_m\rho_m+\omega_{\Lambda}\rho_{\Lambda}).
\end{equation} 
The over dot represents the derivative with respect to time. Combining equations (1) and (2), we get,
\begin{equation}
\dot{H} + 4\pi G(1+\omega_m)\rho_m + 4\pi G(1+\omega_{\Lambda})\rho{\Lambda}= 0.
\end{equation}
Assuming the equation of state for the vacuum energy density as $\omega_{\Lambda}=-1$, we obtain,
\begin{equation}
\dot{H} + 4\pi G(1+\omega_m)\rho_m = 0.
\end{equation}
Using equation (1), the above equation takes the form,
\begin{equation}
\label{eqn:H123}
\dot{H} + \frac{3}{2}(1+\omega_m)H^2 = 4\pi G(1+\omega_m)\rho_{\Lambda}.
\end{equation}
The Renormalisation group approach allows us to consider vacuum energy density as a dynamical quantity \cite{shapiro2005running}. In the cosmic scenario, the vacuum energy's dynamical nature is inherited from its dependence on the apt cosmic variable, the Hubble parameter, $H(t).$ The general covariance of the corresponding effective action restricts the power of the cosmic variable to appear in the dark energy density to an even number. In this model, we assume the vacuum energy density depends on $H^2$ and $H^4.$ The general form of the vacuum energy density can be written as \cite{Sola:2015rra},
\begin{equation}
\label{eqn:rho2}
\rho_{\Lambda}(H) = \frac{3}{8\pi G}\left(c_0 + \nu H^2 + \frac{H^4}{H_I^2}\right).
\end{equation}
In this equation, $c_0$ is a bare cosmological constant that dominates at low energy condition like the one persisted during the Universe's current epoch and the Hubble parameter is close to its present value, $H_0$. The successive terms will give the running status to this density. The parameter $\nu$ is a dimensionless coefficient and is an analog of the $\beta$ function coefficient appearing in the effective action of quantum field theory in curved space time. The last term proportional to $H^4$ has got relevance only in the early epoch of the Universe, the inflationary stage. The constant term $H_I$ is thus equivalent to the constant Hubble parameter during the early inflationary epoch of the Universe. The third term doesn't contain any additional parameter like $\nu$ as in the second term, since any additional parameter which can be proposed will naturally be absorbed into $H_I.$ This equation of the vacuum energy density will give insight into the connection between the early inflation and the late acceleration.      
Substituting equation (\ref{eqn:rho2}) in to (\ref{eqn:H123}), we obtain,
\begin{equation}
\label{eqn:h1}
\dot{H} = \frac{3}{2}(1+\omega_m)c_0 - \frac{3}{2}(1+\omega_m)(1-\nu)H^2+\frac{3}{2}(1+\omega_m)\frac{H^4}{H_I^2}.
\end{equation}
By changing variable from time to scale factor, the equation become easier to handle, thus we have, 
\begin{equation}
aH\frac{dH}{da} = \frac{3}{2}(1+\omega_m)c_0 - \frac{3}{2}(1+\omega_m)(1-\nu)H^2+\frac{3}{2}(1+\omega_m)\frac{H^4}{H_I^2}.
\end{equation} 
On integrating, we obtain Hubble parameter as,
\begin{equation}
\label{eqn:H2}
H(a)  = (1-\nu)^\frac{1}{2}H_I\left(\frac{H_0^2- \frac{C_0}{1-\nu} + \frac{C_0}{1-\nu} a^{3(1+\omega_m)(1-\nu)}}{H_0^2 - \frac{C_0}{1-\nu} + (1-\nu)H_I^2 a^{3(1+\omega_m)(1-\nu)}}\right)^\frac{1}{2}.
\end{equation}
This 
solution
represents the smooth
evolution history of the Universe 
from the early inflationary phase up to the end de Sitter epoch.
In the limit $a\rightarrow 0$, corresponding to the very early epoch, the Hubble parameter would attain a constant value, 
$H \sim (1-\nu)^\frac{1}{2}H_I$ and is corresponding to the early 
inflationary phase of the Universe. 
In the extreme future limit of $a \to \infty,$ the term $(H_0^2 - c_0/(1-\nu))$ in the denominator and the numerator will be negligible, as a result the 
Hubble parameter takes the form, $H 
\sim \left(\frac{c_0}{1-\nu}\right)^{(1/2)}$, 
and it corresponds to the end de Sitter epoch. Between these two accelerating epochs,
the Universe has undergone two transient epochs, first the radiation dominated, and then the matter dominated phase, at each phases the respective Hubble parameters 
will vary as the Universe expands.
\par
To explicitly bring the transient epochs, we use the appropriate equation of state. Due to the small value of the bare constant $c_0$ (which is equivalent to the late cosmological constant responsible for the 
late acceleration), it has no relevance in the early stage. 
Then, on omitting terms containing $c_0,$ the Hubble parameter in 
equation(\ref{eqn:H2}) will takes the form,
\begin{equation}
\label{eqn:H4}
H  \to  (1-\nu)^{1/2} H_I\left(\frac{H_0^2 }{H_0^2 + (1-\nu)H_I^2 a^{3(1+\omega_m)(1-\nu)}}\right)^\frac{1}{2}.
\end{equation}
Then in the very early epoch 
Hubble parameter assumes the form
$H \sim \sqrt{(1-\nu)}H_I$
corresponding to the inflationary epoch. 
After inflation, the Universe would have expanded many times large
so that the Hubble parameter
evolves as equation (\ref{eqn:H4}) and the Universe is in the radiation
dominated era, with equation of state, $\omega_m= 1/3.$ 
\par
We will now consider the next prominent epochs, the matter dominated era and late accelerated epoch, 
during which the most relevant constant is $c_0$ rather than $H_I.$ Thus by putting $\omega_m = 0$ in equation (\ref{eqn:H2}), we obtain,
\begin{equation}
\label{eqn:H10}
H =\left[\frac{H_0^2-\frac{c_0}{1-\nu} + (\frac{c_0}{1-\nu})a^{3(1-\nu)}}{\left(\frac{H_0^2-\frac{c_0}{1-\nu}}{(1-\nu)H_I^2}\right) + a^{3(1-\nu)}}\right]^\frac{1}{2}.
%H = H_0^2  a^{-\frac{3}{2}(1+\omega_m)(1-\nu)}.
%H = H_I\left(\frac{H_0^2 + C_0 a^4}{H_0^2 + H_I^2 a^4}\right)^\frac{1}{2}
\end{equation}
During this period, $\left(\frac{H_0^2-\frac{c_0}{1-\nu}}{(1-\nu)H_I^2}\right) << a^{3(1-\nu)}.$ The term involving $H_I^2$ in the denominator of the above equation is not relevant  
so the equation (\ref{eqn:H10}) reduces to
\begin{equation}
\label{eqn:H11}
H = \left[\left(H_0^2-\frac{c_0}{1-\nu}\right) a^{-3(1-\nu)} + \left(\frac{c_0}{1-\nu}\right)\right]^\frac{1}{2}
\end{equation}
In the asymptotic limit  
$a \to \infty$ 
which 
the Hubble parameter become, $H \sim \left(\frac{c_0}{1-\nu}\right)^\frac{1}{2},$ which is corresponding to the end de Sitter epoch. For the period prior to this, 
the matter dominated epoch, the first term in the RHS of the above equation become the dominant one, and the result is a 
decelerated expansion. This guarantees the transition into the late accelerating epoch from a matter dominated decelerated epoch.
The consistency of this form of the Hubble parameter can be checked by assuming $a=a_0=1$, corresponding to the current epoch and it will result in to 
$H = H_0, $ representing the Hubble parameter value in the present time.
\par
The energy densities at these two successive early epochs can be obtained from the simple Friedman equation $\rho \sim 3H^2$ respectively. The equation (\ref{eqn:H4}) can be re-written as 
\begin{equation}
\label{eqn:H15}
H  \to  H_I\left(\frac{(1-\nu)H_0^2 }{H_0^2 + (1-\nu)H_I^2 a^{3(1+\omega_m)(1-\nu)}}\right)^\frac{1}{2}.
\end{equation}
After inflation the Universe will attain size many times large compared to pre-inflationary period. Hence the scale factor $a$ would be relatively very large in post inflationary era compared to the prior inflationary phase. Hence it is possible to take a local condition, 
the above equation can be suitable approximated, by considering the transition from early inflation to the consequent radiation dominated epoch. This leads to the standard form of the Hubble parameter as,
\begin{equation}
\label{eqn:H06}
H\rightarrow H_0 a^{-2(1-\nu)}.
\end{equation} 
This otherwise implies that the radiation density will evolve as, $\rho_{\gamma} \sim H_0^2 a^{-4(1-\nu)}.$
\par
Now, let us consider the evolution of the energy densities during these early period of inflation-radiation epochs. 
	The
	vacuum energy density,  
	%is given by equation (7). Sinc the 
	by neglecting $c_0$ from equation (\ref{eqn:rho2}) which is not relevant in the initial de-Sitter to radiation dominated phase, and substituting for $H$ using equation (\ref{eqn:H15}),  can be expressed as
\begin{equation}
\label{eqn:rho4}
\rho_{\Lambda}(H) = \rho_{\Lambda_i}\left[\frac{1+\nu(1-\nu)\frac{H_I^2}{H_0^2}a^{4(1-\nu)}}{\left(1+(1-\nu)\frac{H_I^2}{H_0^2}a^{4(1-\nu)}\right)^2}\right].
\end{equation}
Where $\rho_{\Lambda_i} = \frac{3(1-\nu)H_I^2}{8\pi G}$ is the static vacuum energy density corresponding to the initial de Sitter phase of the Universe. That is as $a\rightarrow 0$, $\rho_{\Lambda}(H)\rightarrow \rho_{\Lambda_i}$.
	During the transition form the early inflation to the radiation dominated epoch, the Hubble parameter satisfies the general rule, $H^2=\frac{8\pi G}{3} (\rho_{\Lambda} + \rho_{\gamma}).$
	%The radiation density is given by equation (5). 
	Substituting equation(\ref{eqn:H15}) and (\ref{eqn:rho4}) in (\ref{eqn:h01}) we get the radiation density as,
	\begin{equation}
	\label{eqn:rhogamma}
	\rho_{\gamma}(H) = \rho_{\Lambda_i}\left[\frac{(1-\nu)^2\frac{H_I^2}{H_0^2}a^{4(1-\nu)}}{\left(1+(1-\nu)\frac{H_I^2}{H_0^2}a^{4(1-\nu)}\right)^2}\right].
	\end{equation}
	From the above equation, it is clear that as $a\rightarrow 0$, $\rho_{\gamma}\rightarrow 0$, that is the radiation density is negligible in  the very early phase of the Universe where the vacuum energy is the only dominant component.
\par
Now, let us obtain the analytical expression for vacuum energy density and matter density in the late epoch. The
	vacuum energy density, by neglecting $H^4$ from equation (\ref{eqn:rho2}) which is not relevant in the matter dominated to late de-Sitter phase, and substituting for $H$ using equation (\ref{eqn:H11}),  can be expressed as
	\begin{equation}
	\label{eqn:H14}
	\rho_{\Lambda}(H) = \rho_{\Lambda_0} + \frac{\nu}{1-\nu}\rho_{m_0}\left(a^{-3(1-\nu)}-1\right),
	\end{equation}
	where  $\rho_{m_0} = \frac{3}{8\pi G}\left[\left((1-\nu)H_0^2 - C_0\right)\right]$ and $\rho_{\Lambda_0} = \frac{3H_0^2}{8\pi G} - \rho_{m_0}$ are the present values of the matter and vacuum energy density respectively.
	The present value of the vacuum energy, by substituting $a=1$ in the equation (\ref{eqn:H14}), can be written as,
	$\rho_{\Lambda}(H) = \rho_{\Lambda_0}$.
	In the limit $a\rightarrow\infty$, $\rho_{\Lambda}(H)\rightarrow\rho_{\Lambda_0} - (\frac{\nu}{1-\nu})\rho_{m_0}$.
	This is the vacuum energy density corresponds to the final de Sitter phase of the Universe.
	During the transition from the matter dominated to late accelerating epoch, the Hubble parameter satisfies the general rule, $H^2=\frac{8\pi G}{3} (\rho_{\Lambda} + \rho_{m}).$
	%The radiation density is given by equation (5). 
	Substituting equation(\ref{eqn:H11}) and (\ref{eqn:H14}) in (\ref{eqn:h01}), we get the matter density as,
	\begin{equation}
	\label{eqn:H17}
	\rho_{m}(H) = \rho_{m_0}a^{-3(1-\nu)}.
	\end{equation}
	From the above equation, it is clear that as $a\rightarrow \infty$, $\rho_{m}\rightarrow 0$, that is the matter density is negligibly small in  the final de Sitter phase of the Universe where the vacuum energy is the only dominant component. The present matter, by putting $a=1$ in the equation(\ref{eqn:H17}), can be obtained as $\rho_m(H) = \rho_m^0$.
\section{Finite Boundary for the de Sitter Epochs}
Our Universe is evolving between  two asymptotically de Sitter phases, dominated by vacuum energy density. In between we have a radiation 
dominated phase as well as matter dominated phase \cite{padmanabhan2013cosmin, padmanabhan2014cosmological}. During the transient period, the Universe expands about a factor of $10^{28}$
in the radiation dominated phase, 
%but the Universe 
and expands only about a factor of $10^4$ in the matter dominated phase \cite{Padmanabhan_2012}. So we ignore the matter dominated phase for the ongoing discussion 
to establish the connection between two de Sitter phases.
%Mathematically the two de Sitter phases can last forever. However there should exist physical cut-off length scales 
%in both these accelerated epochs, which make the region 
%of relevance 
%to us be finite. 
The length scale over which the physical processes operate coherently in an expanding Universe is the Hubble radius, $d_H \sim H^{-1},$ and is proportional to the
cosmic time (t) \cite{padmanabhan2013solution}. 
Consider a perturbation generated in the early inflationary period, at some given wavelength scale, $\lambda.$ This will be stretched
with the expansion of the Universe as $\lambda \propto a(t).$

\begin{figure}[h]
	\centering
	\includegraphics[width=8cm,height=6cm]{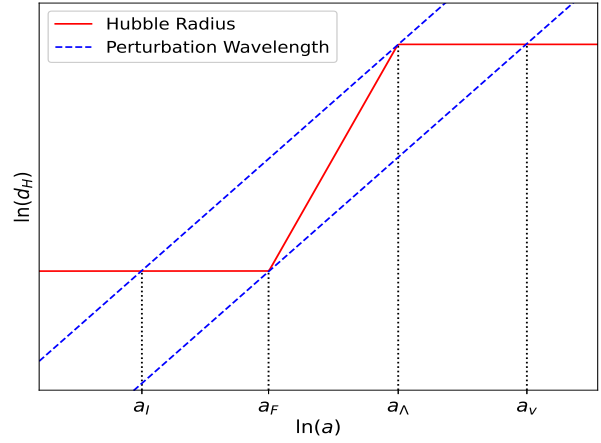}
	\caption{Hubble radius (red solid curve) and wavelength of primordial perturbation (blue dashed lines) are plotted  against the scale factor in logarithmic scale.}
	\label{fig:dH}
\end{figure}

The evolution of Hubble radius with the scale factor in logarithmic scale is plotted in (Fig. \ref{fig:dH}). 
The length $d_H$ will be constant during both the early inflationary
epoch and the end de Sitter epoch. The initial inflationary phase ends at $a=a_F$ and is followed by radiation
and matter dominated phases. These proceeds to another de Sitter phase of late time accelerated
expansion for $a> a_{\Lambda}.$
Mathematically the two de Sitter phases can last forever. However there should exist physical cut-off length scales 
in both these accelerated epochs, which make the region  
to us be finite. Let us obtain $a_F$ the scale factor corresponding to which the energy densities of vacuum and radiation are equal, and after that, the 
radiation density will take over the vacuum density.
%Neglecting $c_0$ in equation (\ref{eqn:rho2}), which is not relevant in the initial de-Sitter epoch,
The vacuum density in the early inflationary phase is given by equation (\ref{eqn:rho4}). After the inflationary period, 
the Universe eventually make transition to radiation dominated phase, the corresponding radiation density can be expressed as in equation (\ref{eqn:rhogamma}).
Equating the two densities at $a=a_F$, we get,
\begin{equation}
\label{eqn:H50}
\rho_{\Lambda_i}\left(\frac{1+\nu(1-\nu)\frac{H_I^2}{H_0^2}a_F^{4(1-\nu)}}{1+(1-\nu)\frac{H_I^2}{H_0^2}a_F^{4(1-\nu)}}\right) = \rho_{\Lambda_i}\left(\frac{(1-\nu)^2\frac{H_I^2}{H_0^2}a_F^{4(1-\nu)}}{1+(1-\nu)\frac{H_I^2}{H_0^2}a_F^{4(1-\nu)}}\right),
\end{equation}
on simplification we get,
% \begin{equation}
% 	\label{eqn:H49}
% 	1+\nu(1-\nu)\frac{H_I^2}{H_0^2}a_F^{4(1-\nu)} = (1-\nu)^2\frac{H_I^2}{H_0^2}a_F^{4(1-\nu)}
% \end{equation}
% Rearranging the above equation, we obtain the scale factor at which the Universe make transition from inflationary phase to radiation dominated phase as,
\begin{equation}
\label{eqn:H13}
a_F = \left(\frac{H_0^2}{(1-2\nu)(1-\nu)H_I^2}\right)^{\frac{1}{4(1-\nu)}}.
\end{equation}
We 
%can 
%also calculate the 
obtain the Hubble radius corresponding to  
%transit from inflationary period to radiation dominated epoch  by substituting the above obtained expression for 
$a_F,$ 
%in 
using equation (\ref{eqn:H4}) as,
\begin{equation}
\label{eqn:H01}
d_{H_F} =\left(\frac{2}{(1-2\nu)H_I^2}\right)^{\frac{1}{2}}.
\end{equation}
Now let us go for the next important scale factor, $a_{\Lambda},$ corresponding to the switch over 
%from the matter dominated epoch 
to the late de Sitter epoch.
The energy density of radiation in the late Universe will evolves as,
\begin{equation}
\label{eqn:H02}
\rho_{\gamma}(a)=\rho_{\gamma}^0a^{-4(1-\nu)}.
\end{equation}
During this late period, the vacuum energy density can be expressed using the expression (\ref{eqn:H14}).
At $a=a_{\Lambda}$ these two densities are equal, that is,
%can be determined by equating the matter energy density and vacuum density,
\begin{equation}
\label{eqn:H03}
\rho_{\gamma}^0a^{-4(1-\nu)} = \rho_{\Lambda}^0+\frac{\nu}{1-\nu}\rho_{\gamma}^0(a^{-4(1-\nu)}-1).  
\end{equation}
Rearranging the above equation give rise to the scale factor $a_{\Lambda}$, as 
%corresponding to which Universe make transition from radiation dominated phase to late accelerating phase, that can be expressed as, 
\begin{equation}
\label{eqn:H18}
a_{\Lambda} = \left(\frac{c_0}{(1-2\nu)[(1-\nu)H_0^2-c_0]}\right)^{\frac{-1}{4(1-\nu)}}.
\end{equation}
The Hubble radius corresponding to the transit from radiation dominated epoch to late accelerating epoch can be obtained by substituting the above 
obtained expression of $a_{\Lambda}$ in equation (\ref{eqn:H11}), as
\begin{equation}
\label{eqn:H27}
d_{H_{\Lambda}} = \left(\frac{1-2\nu}{2c_0}\right)^{\frac{1}{2}}.
\end{equation}

The physical cut-off to both early inflation and late de Sitter epochs can be obtained as follows. During the early inflation, the Hubble radius $d_H$ will 
remain a constant due to prevailing constancy of $H$ during this epoch. On the other hand, the perturbation 
%of a given 
wavelength, $\lambda,$ 
%generated due to 
%the quantum fluctuation in the inflationary epoch, 
grows exponentially, owing to the exponential increase of the scale factor (i.e., the wave will stretch in proportion to 
the scale factor). As a result, thesse
%given 
perturbations will soon leave the Hubble radius.
%say at point A. 
Once the Universe switched over to the radiation dominated era, 
the Hubble radius grows in proportion to $a^2$ ($d_H \propto a^2$), but the perturbation length will still grow in proportion to $a$ only (i.e. $\lambda \propto a$). 
If there would be no late accelrating epoch, then all the perturbation once left the Hubble sphere are eventually re-enter the Hubble sphere in the future course 
of evolution. 
%Consequently, 
But due to the late accelerated epoch, 
%the given perturbation may re-enter the Hubble radius at some future instant, say at B. T
there exists some perturbation, having wavelength beyond some critical value, 
will never enter the Hubble radius.
%once they left during the inflationary epoch. 
This critical length will 
%be proportional to 
naturally causes 
%the 
a finite extension to the early 
de Sitter epoch. The question is how to find this length.

%The physical cut-off to both early inflation and late de Sitter epochs can be obtained as follows. During the early inflation the, the Hubble radius $d_H$ will almost remains a constant, due to prevailing constancy of $H$ during this epoch. On the other hand any perturbation, which is generated due to the quantum fluctuation in the inflationary epoch, of given wavelength, $\lambda$ grows exponentially, owing to the exponential increase of the scale factor (i.e. wave will stretch in proportion to the scale factor). As a result the given perturbation will leave the Hubble radius, say at the point A. Once the Universe switched over to the radiation dominated era, the Hubble radius grow in proportion to $a^2$ ($d_H \propto a^2$).
%But the perturbation length will still grow in proportion to $a$ only (i.e. $\lambda \propto a$). Consequently the given perturbation may would re-enter the Hubble radius at some future instant, say at B. But there exists some perturbation, having wavelength beyond some critical value, will never enter the Hubble radius one they left during the inflationary epoch. This critical length will be proportional to the finite extension of the early de Sitter epoch. The question is how to find this length.

Let us now consider the plot of the $\ln d_H$ versus $\ln a.$. 
The point corresponding to $a_{\Lambda}$ i.e $(\ln a_{\Lambda}, \ln d_{H_{\Lambda}})$ marks the entry of the 
Universe into the late de Sitter epoch. The slope of the tangent to the 
%plotted 
curve at $a_{\Lambda}$, 
%this point 
%corresponding to $a_{\Lambda}$ 
can be obtained as,
%using the relation,
\begin{equation}
\label{eqn:H21}
m=\left(\frac{dln(d_H)}{dln(a)}\right)_{a=a_{\Lambda}}=\frac{a_{\Lambda}}{d_{H_{\Lambda}}}\left(\frac{d(d_H)}{da}\right)_{a=a_{\Lambda}}.
\end{equation}
On detailed calculation this leads to,
\begin{equation}
\label{eqn:H22}
m= \frac{2(1-\nu)(H_0^2-\frac{C_0}{1-\nu})a_{\Lambda}^{-4(1-\nu)}}{(H_0^2-\frac{C_0}{1-\nu})a_{\Lambda}^{-4(1-\nu)}+\frac{C_0}{1-\nu}}.
\end{equation}   
On extending this 
%The 
tangent, it can be found to
%^will 
intersect the flat portion corresponding to the early inflation, at the point $a_I$ i.e.$(\ln a_I, \ln d_{H_I}).$ On 
substituting $a_{\Lambda}^{-4(1-\nu)}$ form equation (\ref{eqn:H18}), it can be verified that the slope, $m=1.$ The pertubation 
wavelength, $\lambda \propto a,$ also have the same slope in this plane. Then it can be concluded that, perturbations leaving the Hubble sphere from beyond 
the location $a_I$ during the early inflation, will never re-enter the Hubble sphere at any time in the future. This means that any region beyond $a_I$ in the inflationary period 
would have no effect on our region, hence it can be take as a boudary which limits the past extend of inflation.
%
% which can be taken as the finite boundary of the inflation in 
% the past. Any perturbation leaving the Hubble sphere beyond $a_I$ will never eneter the Hubble sphere again. 
% Equation (\ref{eqn:H22}) actually represents the slope of the straight line connecting the points $(\ln a_I, \ln d_{H_I})$ and ($\ln a_{\Lambda}, \ln d_{H_{\Lambda}}).$ 
Following the standard equation of the straight line, we can now deduce that,
\begin{equation}
\label{eqn:H23}
a_I = \left(\frac{d_{H_I}}{d_{H_{\Lambda}}}\right)a_{\Lambda},
\end{equation}
where $d_{H_I}$ is the Hubble radius corresponding to the early inflationary phase and it can be expressed as
\begin{equation}
\label{eqn:H45}
d_{H_I} = \frac{1}{(1-\nu)^{\frac{1}{2}}H_I}.
\end{equation}

We will now proceed to  calculate the finite boundary of the end de Sitter epoch. The slope of the tangent at the point $(\ln a_F, \ln d_{H_F})$ as,
%The slope of the tangent to the curve at $a=a_F$ is expressed as
\begin{equation}
\label{eqn:H19}
n=\left(\frac{d\ln(d_H)}{d\ln(a)}\right)_{a=a_F}=\frac{a_F}{d_{H_F}}\left(\frac{d(d_H)}{da}\right)_{a=a_F},
\end{equation}
the detailed form is,
\begin{equation}
\label{eqn:H20}
n= \frac{2(1-\nu)^2\frac{H_I^2}{H_0^2}a_F^{4(1-\nu)}}{1+(1-\nu)\frac{H_I^2}{H_0^2}a_F^{4(1-\nu)}}.
\end{equation}
The tangent will intersect the flat portion corresponding to the late de Sitter phase, at the point $a_v$ (i.e.$(\ln a_v, \ln d_{H_v})),$ 
Substituting for $a_F$ for the previous results, it can be shown that, the slope $n=1.$ Posing similar arguements as in the previous case, the points $a_v,$
%which 
can be taken as the finite boundary of the late de Sitter phase in the future.
%\par The slope of the tangent to the curve at $a=a_F$ is expressed as
%\begin{equation}
%\label{eqn:H21}
%m=\left(\frac{dln(r)}{dln(a)}\right)_{a=a_{\Lambda}}=\frac{a_{\Lambda}}{r_{\Lambda}}\left(\frac{dr}{da}\right)_{a=a_{\Lambda}}
%\end{equation}
%The slope calculated using the above equation can be written as,
%\begin{equation}
%\label{eqn:H22}
%n= \frac{\frac{3}{2}(1-\nu)(H_0^2-\frac{C_0}{1-\nu})a_{\Lambda}^{-3(1-\nu)}}{(H_0^2-\frac{C_0}{1-\nu})a_{\Lambda}^{-3(1-\nu)}+\frac{C_0}{1-\nu}}
%\end{equation}
Equation (\ref{eqn:H20}) actually represents the slope of the straight line connecting the points $(\ln a_F, \ln d_{H_F})$ and ($\ln a_{v}, \ln d_{H_{v}}).$ 
Following the standard equation of the straight line, we can now deduce that,
\begin{equation}
\label{eqn:H24}
a_v = \left(\frac{d_{H_v}}{d_{H_F}}\right)a_{F},
\end{equation}
where $d_{H_v}$ is the Hubble radius corresponding to the late de-Sitter phase and can be expressed as 
\begin{equation}
\label{eqn:H29}
d_{H_{v}}=\left(\frac{1-\nu}{c_0}\right)^{\frac{1}{2}}.
\end{equation}
From equations (\ref{eqn:H23}) and (\ref{eqn:H24}), it 
%these it 
follows that,
\begin{equation}
\label{eqn:H30}
\frac{a_I}{a_F} = \left(\frac{d_{H_I}}{d_{H_{\Lambda}}}\right)  \left(\frac{d_{H_v}}{d_{H_F}}\right) \frac{a_{\Lambda}}{a_v}.
\end{equation}
Substituting the equations (\ref{eqn:H13}), (\ref{eqn:H15}), (\ref{eqn:H18}), (\ref{eqn:H27}), (\ref{eqn:H23}), (\ref{eqn:H45}) in equation (\ref{eqn:H30}), we obtain,
\begin{equation}
\label{eqn:H32}
\frac{a_F}{a_I} = \frac{a_{v}}{a_{\Lambda}}.
\end{equation}
The interesting conclusion from this is that the portions $a_I-a_F$ and $a_{\Lambda}-a_v$ are equal. 
% % During the interim radiation phase the 
% % Hubble radius is, $d_H \propto a^2.$ Since the blue connecting the points corresponding to $A_I$ and $a_{\Lambda}$ is of unit slope, then the red line connecting the 
% % points corresponds to $a_F$ and $a_{\Lambda}$ is line of slope 2. Then we can generally conclude that,
% % \begin{equation}
% %  \frac{a_F}{a_I} = \frac{a_{v}}{a_{\Lambda}} = \frac{a_{\Lambda}}{a_F}
% % \end{equation}
% % We have ignored the matter dominated phase for this calculation. More precise calculation including matter slightly changes the diagram (\ref{fig:dH}). But we can 
% % definitely expect 
% % %ensure 
% % that there exist a definite relationship between physically relevant early inflationary phase and late de Sitter phase. The above results otherwise implies that,
% % the number of perturbation modes leaving the Hubble region between $a_I$ and $a_F$ will be equal to the the number of modes entering the region between $a_{\Lambda}$ and $a_v.$
%\section{CosMIn: The conserved number}
%We have seen in the previous section that our Universe has three distinct phases of evolution: 
There is a middle portion in the pot corresponds to $a_F - a_{\Lambda},$ which represents the interim radiation/matter phase. Let us now check the relative 
duration of this interim phase and contrast it the duration of the first and last de Sitter epochs. From equation (\ref{eqn:H06}), it is clear that the slanted portion of the curve representing the Hubble radius in (Fig. \ref{fig:dH}) has a slope $2(1-\nu)$. Assuming $d_{H_F}$ almost equal to $d_{H_I}$, we can express the slope as,
\begin{equation}
\label{eqn:H07}
2(1-\nu) = \frac{\ln\left(\frac{d_{H_\Lambda}}{d_{H_I}}\right)}{\ln\left(\frac{a_{\Lambda}}{a_{F}}\right)}.
\end{equation}
Similarly, the slanted portion of the curve representing perturbation wavelength in (Fig. \ref{fig:dH}) has a unit slope. So we can express the slope as,
\begin{equation}
\label{eqn:H08}
1 = \frac{\ln\left(\frac{d_{H_\Lambda}}{d_{H_I}}\right)}{\ln\left(\frac{a_{\Lambda}}{a_{I}}\right)}.
\end{equation}
combining equation (\ref{eqn:H07}) and (\ref{eqn:H08}), we obtain,
\begin{equation}
\label{eqn:H09}
\left(\frac{a_{\Lambda}}{a_F}\right)^{1-\nu} = \frac{a_{F}}{a_I}.
\end{equation}
From the above equation, it is clear that the portions $a_I-a_F$ and $a_{\Lambda}-a_v$ become equal to $a_F-a_{\Lambda}$ only if $\nu = 0$ \cite{Padmanabhan_2012}.
However the parameter $\nu$ is relatively very small, 
hence it can be conclude that all the three intervals are almost equal to each other.

\par
The above proven approximate equality has an interesting significance. It turn out that
%Now, let us consider 
the number of 
perturbation modes leaving the horizon during the 
early phase $a_I - a_F$, 
%that 
will re-enter the horizon during the interval $a_F - a_v $ and will exit the Hubble sphere again during $a_{\Lambda} - a_v.$ 
%the This number will in turn equal to the number of modes leaving the 
% \par
% Now, let us consider number of perturbation modes leaving the horizon during the 
% early phase $a_I - a_F$, that will be equal to the number of modes enter the horizon during $a_F - a_v.$ This number will in turn equal to the number of modes leaving the 
% last de Sitter epoch $a_{\Lambda} - a_v.$ 
% An early inflationary phase that is driven by the vacuum energy density during which the Universe expand very rapidly, the late accelerating phase that is driven by a 
% slowly varying vacuum energy density and a radiation and matter dominated phase in between. There exist a dimensionless conserved number, 
% Cosmic Mode Index (CosMIn) which enumerate the number of modes crossing the Hubble radius during three different phases of the Universe. 
%Now, 
%we have to estimate the number of modes that leave the Hubble radius during the interval $a_I<a<a_F$. We can show that the same number of modes 
%re-enter during $a_F<a<a_{\Lambda}$ and that again exit during $a_{\Lambda}<a<a_v$. 
%The 
A perturbation of wavelength 
%of perturbation 
$\lambda_p = a/k$ (where k is the 
comoving wave number), crosses the Hubble radius when $\lambda_p(a)=H^{-1}$ or equivalently $k=aH(a)$ is satisfied. The modes having comoving wave numbers 
within the domain $(k, k+dk)$ cross the respective Hubble radius during the interval $(a,a+da)$. The number of modes with wave number $dk$ can be written 
as $dN=V_{com}d^3k/(2\pi)^3$ (where $v_{com}$ is the comoving volume).
The number of modes crossing the Hubble radius during the interval $(a_I<a<a_F)$ can be written as
\begin{equation}
N(a_I,a_F) = \int_{a_I}^{a_F}\frac{V_{com}k^2}{2\pi^2}\frac{dk}{da}da =\frac{2}{3\pi}\int_{H_I a_I}^{H_F a_F}\frac{d(Ha)}{Ha} = \frac{2}{3\pi}\ln\left(\frac{H_Fa_F}{H_Ia_I}\right)
\end{equation}
Substituting for $a_I$ form equation (\ref{eqn:H23}), gives 
%in (48), we obtain,
\begin{equation}
\label{eqn:N12}
N(a_I,a_F) =  \frac{2}{3\pi}\ln\left(\frac{H_Fa_F}{H_{\Lambda}a_{\Lambda}}\right) 
%= N(a_{\Lambda},a_F).
\end{equation}
The RHS of the above equation is $N(a_F,a_{\Lambda})$ the number modes crossing the Hubble radius of the radiation dominated era.
Thus 
%we obtained that 
the number of modes leaving the horizon during early inflationary period will re-enter the radiation dominated era. 
%is equal to the number of modes enter the horizon during the interim 
%radiation dominated epoch.
Now substitute for $a_F$ in equation (\ref{eqn:N12}) from equation (\ref{eqn:H24})
%substituting equation (43) in (49), 
we obtain
\begin{equation}
\label{eqn:N13}
N(a_I,a_F) =\frac{2}{3\pi}\ln\left(\frac{H_va_v}{H_{\Lambda}a_{\Lambda}}\right) 
%= N(a_{\Lambda},a_v)
\end{equation}
which in turn equal to  $N(a_{\Lambda},a_v).$
From equation (\ref{eqn:N12}) and equation (\ref{eqn:N13}), we obtain that the number of perturbations modes leaving the Hubble radius of the 
early inflationary phase, subsequently re-enter the radiation pahse and finally exit the end de Sitter epoch corrsponding to the late acceleration, hence
\begin{equation}
\label{eqn:N14}
N(a_I,a_F) = N(a_F,a_{\Lambda}) = N(a_{\Lambda},a_v).
\end{equation}
This otherwise implies that, $\left(\frac{H_Fa_F}{H_{\Lambda}a_{\Lambda}}\right)=\left(\frac{H_va_v}{H_{\Lambda}a_{\Lambda}}\right),$ which 
in turn equal to $\left(\frac{H_Fa_F}{H_Ia_I}\right).$ This equality, however doesn't implies the equality of the intervals in the scale factor,
especially in the case of the interim radiation (or matter) dominated epoch.
%So it seems that this 
The number of modes crossing the Hubble radii in subsequent epochs is thus a conserved quantiy and let it be 
%Here we get an interesting result that the number modes crossing each epoch is same even though the corresponding intervals are different. 
%Hence, this 
%we can conclude 
%confirm our earlier statement that the number of modes that cross the Hubble radius during the three distinct phases of the Universe is a characteristic conserved number 
%CosMIn, 
%can be 
denoted by $N_c$ for our Universe. In the following we evaluate this quantiy.
%Now, let us evaluate $N_c$ explicitly. To do that, 
Consider the radiation dominated epoch. The number of modes crossing the Hubble radius during the interval $a_F < a < a_{\Lambda}$ can be written as,
\begin{equation}
\label{eqn:N15}
N_c = \frac{2}{3\pi}\ln\left(\frac{H_{F}a_{F}}{H_{\Lambda} a_{\Lambda}}\right) = \frac{2}{3\pi}\ln\left(\frac{d_{H_{\Lambda}}a_{F}}{d_{H_{F}} a_{\Lambda}}\right),
\end{equation}
Substituting equation (\ref{eqn:H01}), (\ref{eqn:H13}), (\ref{eqn:H18}) and (\ref{eqn:H27}) in equation (\ref{eqn:N15}), we obtain
\begin{equation}
\begin{split}
\label{eqn:N16}
N_c = \frac{2}{3\pi}\ln\left[\frac{(1-2\nu)H_I}{2\sqrt{c_0}} \left(\frac{\sqrt{c_0}}{(1-2\nu)(1-\nu)H_I\left(\sqrt{1-(\frac{c_0}{(1-\nu)H_0^2})}\right)}\right)^{\frac{1}{2(1-\nu)}}\right]
\end{split}
\end{equation}
an on simplification, we arrive at,
\begin{equation}
\begin{split}
\label{eqn:N17}
N_c \sim \frac{(1-2\nu)}{3\pi(1-\nu)}\ln\left(\frac{(1-2\nu)H_I}{\sqrt{c_0}}\right) - \frac{2}{3\pi}\ln(2) - \frac{1}{3\pi(1-\nu)}\ln(1-\nu)
\end{split}
\end{equation}
The second and third term in  equation (\ref{eqn:N17}) are negligible small compared to first term, hence we obtain
\begin{equation}
\begin{split}
\label{eqn:N18}
N_c \sim \frac{(1-2\nu)}{3\pi(1-\nu)}\ln\left(\frac{(1-2\nu)H_I}{\sqrt{c_0}}\right).
\end{split}
\end{equation}
%Note that, 
Here $H_I$ is the Hubble parameter of the initial de Sitter Universe, and 
%its magnitude is of the order of GUT energy scale, $\approx10^{14}$ GeV. 
%The 
$c_0$ represents the Hubble parameter corresponds to the late de Sitter Universe. 
%, its magnitude is of the order of cosmological constant, $\approx10^{-42}$ GeV. 
The model parameter $\nu \leq 10^{-3}$ and if it assumed to be negligibly small, then $N_c \sim (1/3\pi) \ln\left(\frac{H_I}{\sqrt{c_0}}\right).$ 
Assuming the standard values, $H_I \sim 10^{14}$ GeV and  $\sqrt{c_0} \sim 10^{-42}$ GeV, it can be shown that, 
%Using these very values and with maximum value for the aparameter $\nu,$ an estimate of $N_c$ is made and found that it comes to around 
%rough value of these constants, we obtain, 
$N_c \sim 4\pi.$ Here we have expressed the numerical value by retaining $\pi$ because we have that in the denominator of the arguement of the logarithm.
\section{Conclusions}
In this work, we have derived an analytical solution for the Hubble parameter for the
complete background evolution of the Universe; from the early inflation to late de
Sitter phase using the running vacuum model of the dark energy proposed by Sola
and others. The general form of the Hubble parameter is suitably reduced to the one which corresponds to the subsequent phases in the evolution of the Universe, the early inflation, interim radiation (matter), and end de Sitter epochs through a
late accelerated phase. The Universe, with evolution as shown in figure (\ref{fig:dH}) having
two distinct de Sitter epochs, one during the inflation and the other during the late time acceleration. 
Both of these epochs can be extended indefinitely into the past and future with a constant
Hubble radius. Nevertheless, there are physical processes that limit the physically relevant region of these two epochs.
Since the Hubble radius flattens out when $a > a_{\Lambda}$, the perturbations with wavelengths larger than a critical value will
never re-enter the Hubble radius, which we imply a physical boundary of this de Sitter epoch. This otherwise implies that only those perturbations that left the early inflationary period during $a_I < a < a_F$ are feasible, which in turn determine the early de Sitter's finite boundary epoch.
We found that the resepctive scale factors will satisfy a relation, $(a_F/a_I)=(a_v/a_{\Lambda}).$ 
% The precise description
%of 
The transition between the two de Sitter phases is 
%the standard domain of conventional cosmology
%in which, depending on the dynamics of the matter sector, one will have a 
through a radiation dominated phase,
giving way to a very late time matter-dominated phase. It is, however, evident that 
%in the overall cos-
%mological evolution, 
the matter-dominated epoch is not 
%of 
much significant since it 
%again 
quickly
gives way to the second de Sitter phase dominated by the cosmological constant. Hence we considered only the radiation epoch as the transient phase. On analyzing 
the evolution of this interim epoch, we found the scale factors ratio, which can equate with the previously mentioned ratio is $(a_{\Lambda}/a_F)^{1-\nu}.$ One may 
note at this juncture that the slanted portion in figure (\ref{fig:dH}), which represents the evolution of the interim phase, has a slope of 2$(1-\nu).$

%Note that the
The perturbation modes which exit the Hubble radius during $a_I-a_F$ re-enter the Hubble radius during $a_F-a_{\Lambda}$ and again
exit during $a_{\Lambda}-a_v.$ We obtained this conserved number of modes and is approximately $N_c \sim (1/3\pi) \ln\left(\frac{H_I}{\sqrt{c_0}}\right).$ We expect 
that this constant number may have further insights into the connection between the early inflationary epoch and the end de Sitter epoch, about which more work is needed.
\bibliography{paper_dec12}
\bibliographystyle{ieeetr}
\end{document}